\documentclass[10pt]{iopart} 
\usepackage{epsfig}

\begin{document}

\title[Accelerating networks]{Accelerating networks}
\author{David M.D. Smith$^{1}$}
\author{Jukka-Pekka Onnela$^{1,2}$}
\author{Neil F. Johnson$^{1}$}

\address{$^1$Clarendon Laboratory, Physics Department, Oxford University, Oxford, U.K}
\address{$^2$Laboratory of Computational Engineering, Helsinki University of Technology, Finland}

\begin{abstract}
Evolving out-of-equilibrium networks have been under intense scrutiny recently. In many real-world settings the number of links added per new node is not constant but depends on the time at which the node is introduced in the system. This simple idea gives rise to the concept of accelerating networks, for which we review an existing definition and -- after finding it somewhat constrictive -- offer a new definition. The new definition provided here views network acceleration as a time dependent property of a given system, as opposed to being a property of the specific algorithm applied to grow the network. The defnition also covers both unweighted and weighted networks. As time-stamped network data becomes increasingly available, the proposed measures may be easily carried out on empirical datasets. As a simple case study we apply the concepts to study the evolution of three different instances of Wikipedia, namely, those in English, German, and Japanese, and find that the networks undergo different acceleration regimes in their evolution.
\end{abstract}

\maketitle

\section{Introduction} \label{sec:introduction}
In many real-world networks the rate at which links are added to the network is different from the rate at which nodes are added to it, and this seems to be the case in particular in  ``functionally organized systems whose operation is reliant on the integrated activity of any or all of its component nodes" \cite{mattick:2005}. This results in accelerating or decelerating networks, of which the Internet is one example \cite{internet}. Other examples of such systems include the arrangement of components on an integrated chip, configuration of computers in clusters, structure of integrated production systems, configuration of human social and communication systems, and the organisation of regulatory proteins controlling gene expression in bacteria \cite{mattick:2005}. 

\begin{figure}[htbp]
\begin{center}
\includegraphics[width=0.7\textwidth]{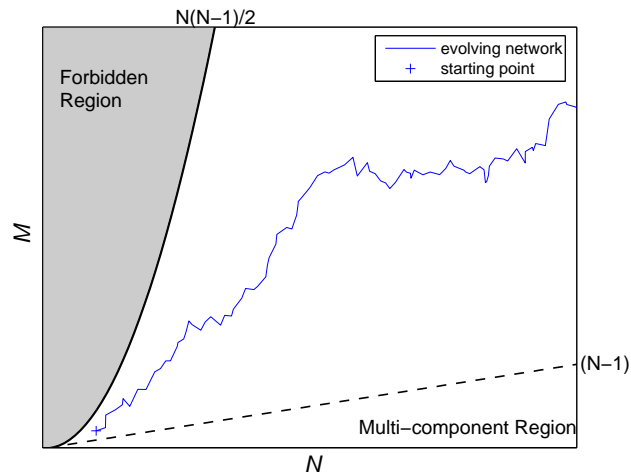}
\caption{ \label{fig:phase} The trajectory of an evolving network might exist anywhere in the allowable region of the phase space corresponding to its number of nodes $N(t)$ and number of links $M(t)$. This might be a network evolving according to some algorithm or an empirically observed network. Whilst it is possible that at any stage in its evolution the network might have many components, the criterion $M(t)<(N(t)-1)$ imposes the condition. }
\end{center} 
\end{figure}

Consider an arbitrary evolving network that may explore any allowed trajectory in its phase space of node number $N(t)$ (number of nodes at time $t$) and link number $M(t)$ (number of links at time $t$). A constructed example of such a network is shown in Fig.~\ref{fig:phase}. In an undirected network of size $N(t)$, the total number of possible links is $N(t)(N(t)-1)/2$ of which $M(t)$ exist at some time $t$. The ratio of the two is described by 

\begin{eqnarray}\label{eqn:ratio1}
q(t)&=&\frac{2~M(t)}{N(t)(N(t)-1)}.
\end{eqnarray} 

In many conventional non-equilibrium evolving (growing) networks this quantity is expected to decrease for large values of $t$. As time-stamped network data becomes increasingly available, we expect most networks to exhibit non-trivial behaviour with respect to their acceleration characteristics. The notion of accelerating networks has been considered before \cite{GM,Sen}, and here we consider the notions of network acceleration as discussed by Gagen and Mattick in \cite{GM}. Although interesting, the definitions in \cite{GM} seem constrictive in that they are applicable only to a specific algorithm, and the concept of acceleration is static in nature, i.e., a given network is considered to be either accelerating, non-accelerating, or decelerating throughout its evolution. Contrary to this, the notion of acceleration in physical systems refers to the nature of the evolution of a system at a specific moment in time, suggesting that acceleration in the context of complex networks should be reconsidered. In addition, it is imperative that network acceleration be measurable for empirical networks. 

We review the model of Gagen and Mattick in Section \ref{sec:GM} and critically examine the proposed concept of network acceleration in Section \ref{sec:approach}. This is followed by a new definition of network acceleration in Section \ref{sec:definition1}. This might be used to describe a network at any stage of its evolution (whether decaying, growing, or neither), and applies to both directed and undirected networks. We then extend the analysis to incorporate weighted links. We demonstrate the use of the new definition in a simple empirical case study in Section \ref{sec:case}, and conclude our discussion in Section \ref{sec:conclusion}.


\section{The model of Gagen and Mattick}\label{sec:GM}

\begin{figure}[htbp]
\begin{center}
\includegraphics[width=0.6\textwidth]{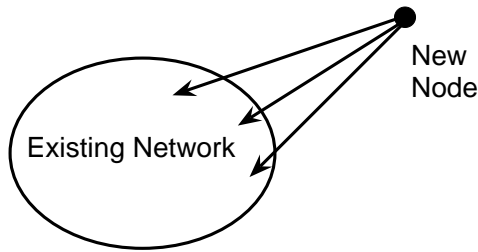}
\caption{ \label{fig:evolvntet2} Evolution of a network. In most out-of-equilibrium network models a new node attaches to the existing network with $m$ links at each time step via some arbitrary attachment mechanism. These links might be directed or undirected, weighted or unweighted.}
\end{center} 
\end{figure}

In the model of Gagen and Mattick (GM-model), as in many other conventional out-of-equilibrium evolving network models, the system evolves by introducing \emph{exactly} one new node at each time step. The key feature of the model is that the new node connects to the existing single-component network through a time-dependent number of links, whereas in most evolving network models \cite{dorogovtsev} the new node attaches to the existing network with a fixed, time-independent number of links $m$, as demonstrated in Fig.~\ref{fig:evolvntet2}.

\begin{figure}[htbp]
\begin{center}
\includegraphics[width=0.6\textwidth]{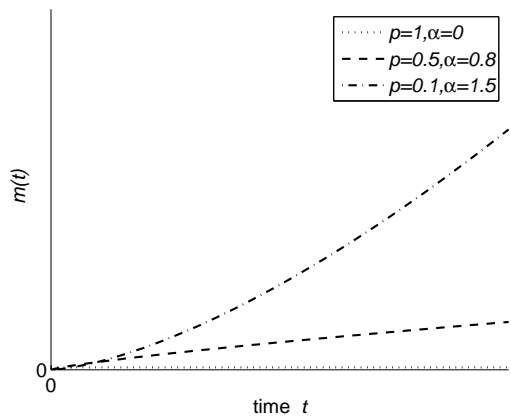}
\caption{ \label{fig:GM1} The evolution of the expected number of links added per new node, $m(t) = p (N(t))^{\alpha}$, within the Gagen and Mattick model of accelerating networks. Three examples comprising non-acceleration ($\alpha=0$), acceleration ($\alpha=0.8$) and hyper-acceleration ($\alpha=1.5$) are displayed \cite{GM}.}
\end{center} 
\end{figure}

Here we shall consider the scenario whereby the new node attaches with undirected binary (unweighted) links. Specifically, we assume that each new node attaches with \emph{on average} $m(t)$ links. Whether or not the number of links added per time step is a stochastic process, the actual evolution of the system will be if a stochastic attachment algorithm (random, preferential, or otherwise) is employed. Clearly, the maximum number of links with which the new node can connect with to the existing network, is equal to the number of nodes within the existing network. Gagen and Mattick actually specify the functional form describing the expected rate of link addition $m(t)$ as
\begin{eqnarray}~\label{eqn:GM}
m(t)&=& p\big(N(t)\big)^\alpha. 
\end{eqnarray}
Here $\alpha$ is an described as an acceleration parameter and $p$ as a probability constant with the constraint  of $0 \le p \le 1$. Gagen and Mattick describe the type of a network  in terms of $\alpha$ as either decelerating ($\alpha < 0$), non-accelerating ($\alpha=0$), accelerating ($0< \alpha<1$), or hyper-accelerating ($\alpha \ge 1$). Some examples are shown in Fig.~\ref{fig:GM1}. We shall revisit such definitions later.

We note that for a non-accelerating network ($\alpha = 0$) within this description, the expected number of new links added is simply $p \in [0,1]$ such that, on average, at most one link may be added per time step. Note also that the hyper-accelerating network has a finite time scale for which it can be applied in that the maximum number of links that can possibly be added to the system is $N(t)$. This sets an upper bound on $m(t)$ after which the network cannot hyper-accelerate, i.e., in order to continue hyper-acceleration, it would have to introduce more than $N(t)$ links per time step, which is impossible without allowing for multiple links between any pair of nodes.
 
We also note also that so far the actual attachment mechanism, random or otherwise, has not been discussed, meaning that the \emph{accelerating nature of the network is simply related to the rate of link addition}. This is an important point to consider because although the expected number of links added for the new node is $m(t)$, the variance in this will differ between microscopic attachment mechanisms. Indeed, the variance could be zero for integer $m(t)$, corresponding to exactly $m(t)$ begin added every time step, or the number of links to add could be drawn from a probability distribution with mean $m(t)$. Even though the number of links to be added might be deterministic, one still needs to specify the algorithm to determine to which nodes within the existing network these links are attached. 

\section{Rethinking the accelerating network}\label{sec:approach}
The notion of network acceleration is applicable to situations whereby the rates of node and link addition are not stochastic. As such, in introducing the concepts key to their understanding, it is useful to concentrate on this conceptually simpler scenario.

\begin{figure}[htbp]
\begin{center}
\includegraphics[width=0.6\textwidth]{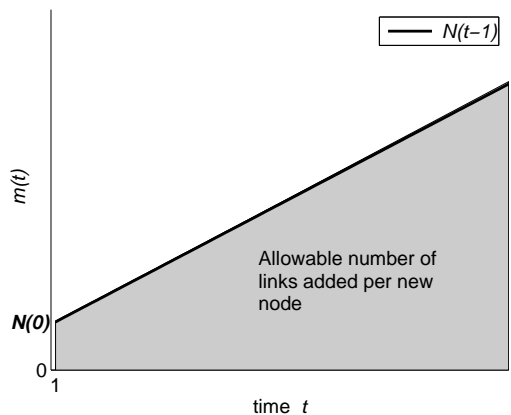}
\caption{ \label{fig:shaded} The allowable number of links (shaded area) added with each new node is capped by the total number of nodes in the existing network (line). Here, as one new node is added per time step, the growth of the network in terms of number of nodes is linear.}
\end{center} 
\end{figure}

Consider some arbitrary evolving network whose evolution we can observe. This might be a realisation of an algorithm or an empirically observed evolving network. At time $t=0$ we have an initial configuration of  $N(0)$ nodes and $M(0)$ links. At each time step, a quantity of new nodes $n(t)$ are added and they are connected to the pre-existing network with some number of binary links $m(t)$. These are both integer quantities. In this scenario no new links are formed between existing nodes within the network, although this feature could be incorporated. For now, we shall assume that $n(t) = 1$ such that exactly one new node is added per time step\footnote{Often, there is little merit in adding more than one node per time step as this might be equivalent to adding one node per time step over a longer period. However, situations might arise, virtual or empirical, in which several nodes are added per time step.}. As such, at time $t=1$ the number of nodes is $N(1)=N(0)+1$. The maximum number of links that could have been introduced on attaching the $N(1)$th node is clearly $N(0)$ as the new node can link to each previous node with at most one link. Similarly, for a node added at time $t$, the total number of nodes is $N(t) = N(0)+t$ and the maximum number of links that could have been used to connect it is $N(t-1)$.  For this process, by which one new node is added per time step, we know that $N(t-1)=N(t)-1$. This is the upper limit for the time-dependent number of added links, $m(t)$, at each time step. This region is depicted Fig.~\ref{fig:shaded}. 

Clearly, the functional form of $m(t)$ could be any function that exists within the allowable region of $m(t)\le (N(t)-1)$ for the addition of one node per time step. 
For the addition of $n(t)$ nodes per time step, the constraint becomes

\begin{eqnarray}\label{eqn:ntconstraint}
\frac{m(t)}{n(t)}&\le&N(t-1).
\end{eqnarray}
One can envisage any number of such functions for the time-dependent number of links added. We note that once the function reaches the constraint, the network can no longer evolve according to such a process.

\begin{figure}[htbp]
\begin{center}
\includegraphics[width=0.49\textwidth]{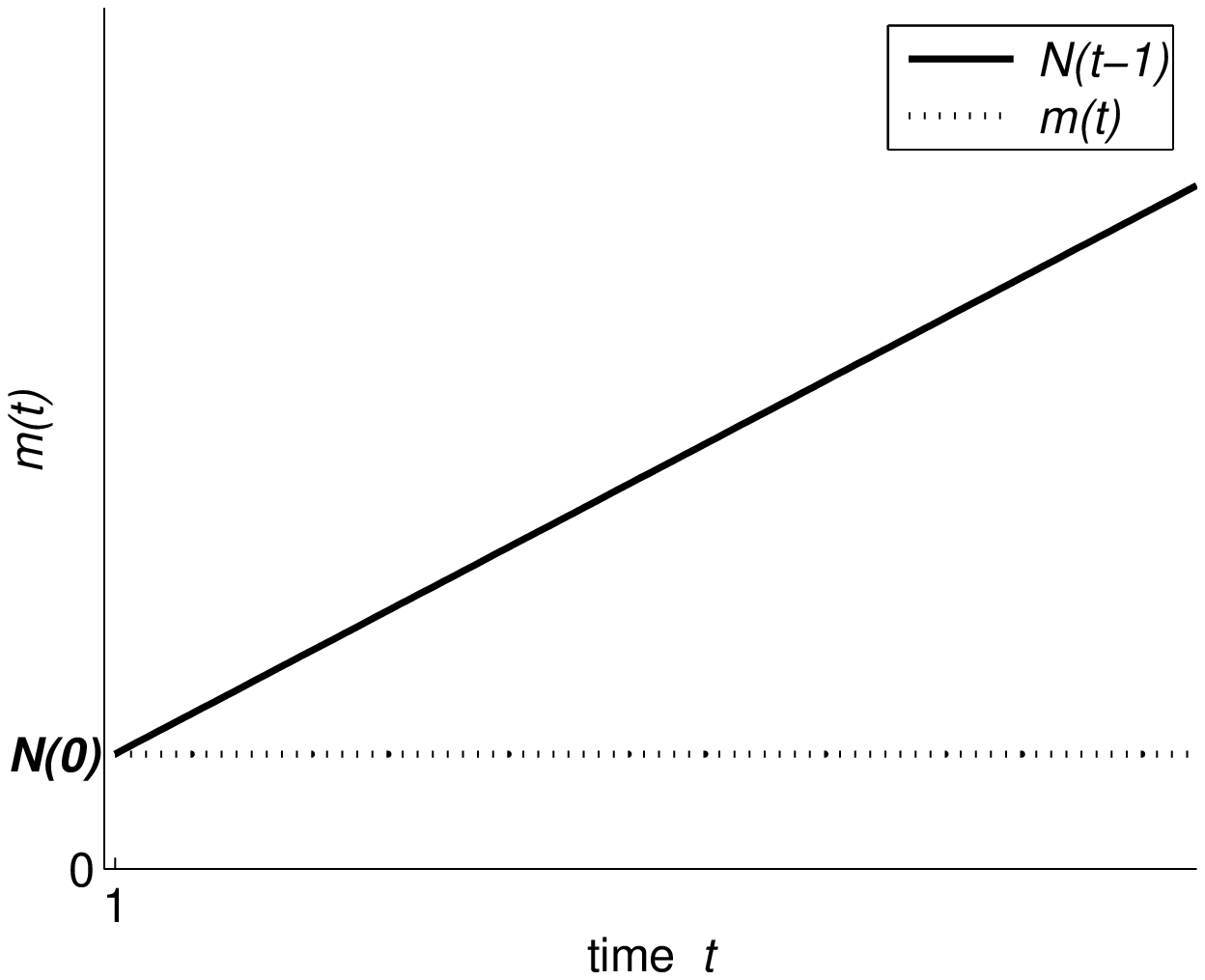}
\includegraphics[width=0.49\textwidth]{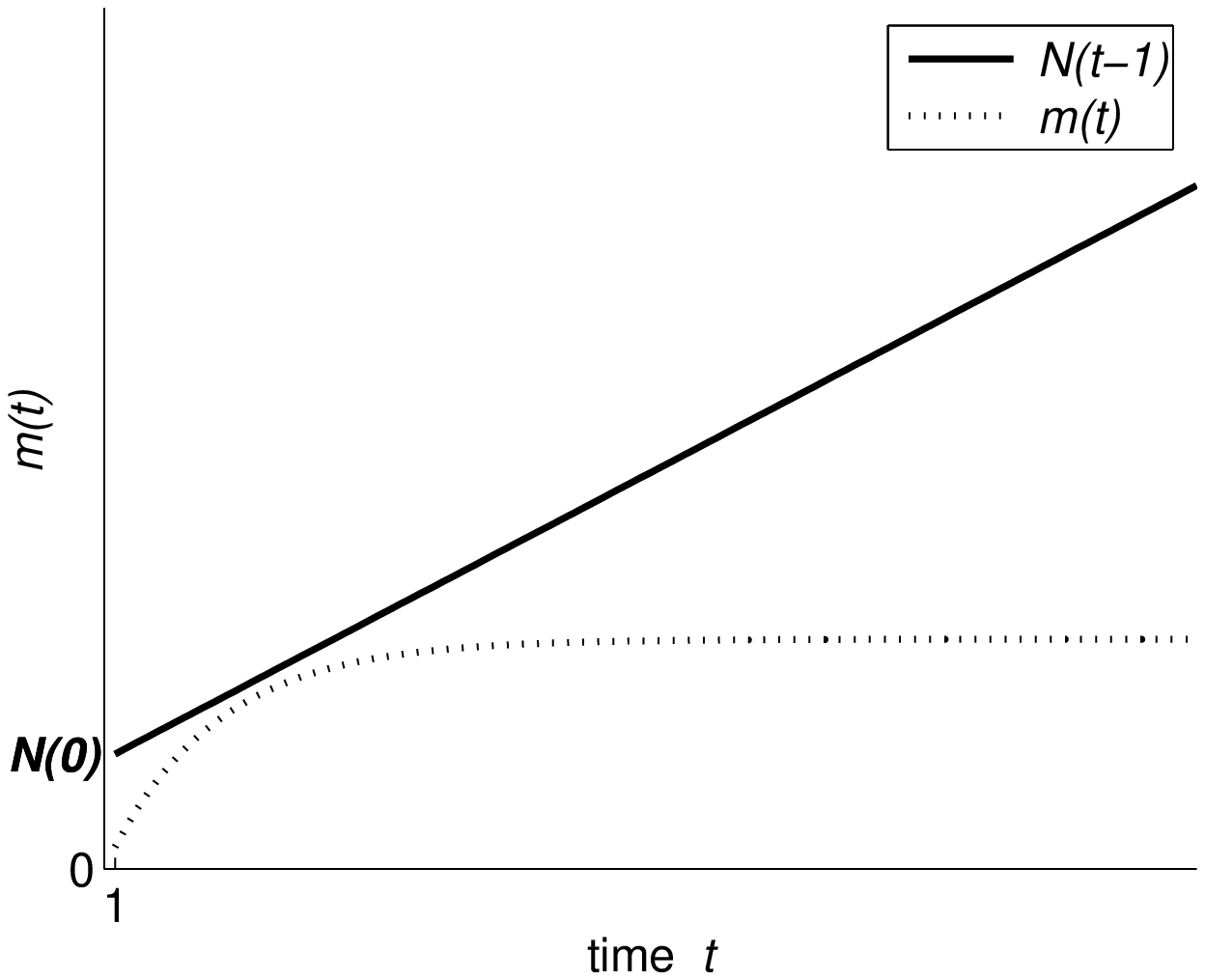}
\caption{ \label{fig:NEWACC} Left: A fairly standard example of a non-accelerating network in which more that one link  is used to link the new node to the existing network. Right: Although this evolving network asymptotes to non-accelerating behaviour, clearly, initially, it is accelerating.}
\end{center} 
\end{figure}
A simple example might be such that a single new node is added with $m(t)= N(0)$ links per time step with $N(0)>1$  as depicted in Fig.~\ref{fig:NEWACC}. This is a non-accelerating network that could not be described by the GM-model that only allows at most one new link per new node for non-accelerating networks, corresponding to $p=1$ and $\alpha=0$ in their model. Another example function is one that is initially increasing but asymptotes to a constant value as depicted in Fig.~\ref{fig:NEWACC}. This might be some empirically observed network growth or the behaviour of some growth algorithm.  Would this be described an accelerating, decelerating or non-accelerating network? Clearly, different regimes of this particular network evolution  might satisfy differing descriptions. As such, we must re-define the accelerating network to encompass this feature.

\section{Defining accelerating networks}\label{sec:definition1}
One might expect that one could identify the regimes of accelerating, non-acceleration and deceleration with relative ease, writing these phenomena in terms of the mean degree $\langle k \rangle = 2M(t) / N(t)$ of the network. Specifically, one might expect that if the addition of a new node via some number of links results in the mean degree of the entire network to increase, the network would be described as accelerating. Likewise, if the mean degree remains constant, the network is not accelerating and, if the mean degree decreases with the addition of new nodes, the network could be considered decelerating. 
\begin{figure}[htbp]
\begin{center}
\includegraphics[width=0.7\textwidth]{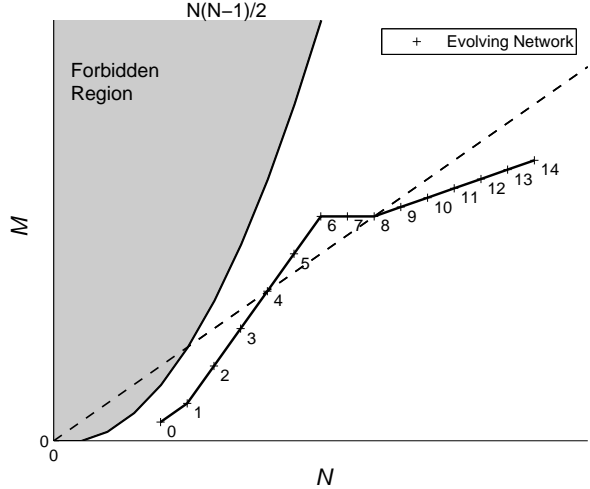}
\caption{ \label{fig:meanplot} A schematic illustration of a network evolving over fourteen time steps in the link and node phase space (solid line). The mean degree of the network is given by $\langle k \rangle = 2M(t) / N(t)$, and the dashed line represents half the mean degree of the system at time step $8$ (and time step $4$). At time step $8$ the network accelerates but the mean degree actually decreases.}
\end{center} 
\end{figure}

Although these ideas are intuitively appealing, one can envisage a scenario whereby a network might be accelerating without increasing its mean degree. This event can occur if a network has undergone rapid deceleration such that the rate of node addition is very low even though it might have been high previously. At some point the ratio of number of links added to the number of nodes added over a time step, $[M(t) - M(t-1)] / [N(t) - N(t-1)]$, might be less than that for the existing network as a whole, $M(t) / N(t)$, thereby decreasing the average degree $\langle k \rangle$ while still constituting network acceleration. This is evident in Fig.~\ref{fig:meanplot}, in which at time step $t=8$ the network accelerates, although the mean degree of the system decreases as the ratio $M(t) / N(t)$ decreases. In order for the mean degree to increase the trajectory would have to exceed the dash line whose gradient represents half the mean degree of the system at time $t=8$.

In order to identify the different regimes of network acceleration, we must relate the rate of increase in the number of links with the rate of addition of new nodes, denoting the rate of link addition and node addition by the approximate derivatives

\begin{eqnarray} \label{eqn:approximate}
m(t) = & \frac{d M(t)}{d t} & \approx M(t) - M(t-1) \nonumber\\
n(t) = & \frac{d N(t)}{d t} & \approx N(t) - N(t-1).
\end{eqnarray}
We can then define the regimes of network acceleration. The important ratio is that of the rate of link addition to the rate of node addition, $m(t)/n(t)$, the evolution of which prescribes a network measure. We define network \emph{acceleration} $a(t)$ as  

\begin{eqnarray} \label{eqn:ratio}
a(t) & \equiv & \frac{d}{d t}\left(\frac{m(t)}{n(t)}\right). 
\end{eqnarray}
We approximate the discrete values with continuous derivatives and define the following three regimes:

\begin{eqnarray} \label{eqn:definitions}
\left\{ \begin{array}{lll} 
a(t) & < 0 & \quad \textrm{decelerating} \nonumber\\
a(t) & = 0 & \quad \textrm{non-accelerating} \nonumber\\
a(t) & > 0 & \quad \textrm{accelerating}.
\end{array} \right. 
\end{eqnarray}
As such, a single evolving network might navigate all regimes. Note that the definition of $a(t)$ allows more than one node to be added per time step. It is interesting to note that within this definition of network acceleration, a decaying network (losing nodes) could accelerate. Also, we note that the definition holds for directed graphs. The above definition alludes to the notion of network \emph{velocity} $v(t)$, which we define as

\begin{eqnarray} \label{eqn:velocity}
v(t) \equiv  \frac{d M(t)}{d N(t)} = \frac{m(t)}{n(t)}.
\end{eqnarray}
This velocity is simply the gradient of the network trajectory in the link-node phase space as in Fig.~\ref{fig:meanplot}.

\subsection{Note on hyper-acceleration}\label{subsec:hyper}
We note the existence of a turning point in the accelerating $a(t)>0$ regime of network evolution.
The acceleration regime $a(t)>n(t)$ cannot be sustained indefinitely as the number of added links per new node would have to exceed the number of existing nodes $N(t)$ which is not possible. As such, this behaviour is deemed {\it hyper-acceleration} if 

\begin{equation} ~\label{eqn:hyper}
a(t) >n(t).
\end{equation}

\begin{figure}[htbp]
\begin{center}
\includegraphics[width=0.7\textwidth]{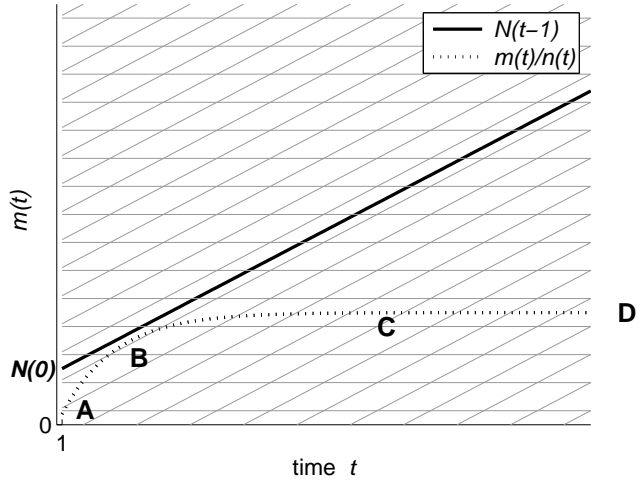}
\caption{ \label{fig:NEWACC3} Superimposing the appropriate contour lines, we observe hyper-acceleration between {\bf A} and {\bf B}, acceleration between {\bf B} and {\bf C} and non-acceleration between {\bf C} and {\bf D}.}
\end{center} 
\end{figure}

\begin{figure}[htbp]
\begin{center}
\includegraphics[width=0.7\textwidth]{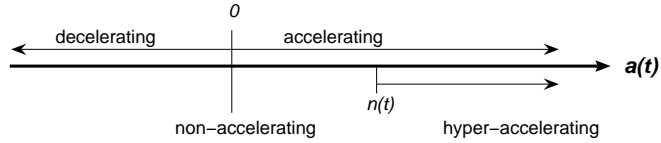}
\caption{ \label{fig:Rdiag} Different regimes of network acceleration.}
\end{center} 
\end{figure}

If we reconsider the function $m(t)$ being an initially increasing function of time asymptoting to a constant value and superimpose the appropriate contour lines, i.e. $y = const$ and $y =n~x+const$, where coefficient $n$ corresponds to the constant rate of node addition $n(t)=n$, one can clearly identify the acceleration regimes for this particular evolving network. This evolution is depicted  in  Fig.~\ref{fig:NEWACC3}. We observe hyper-acceleration between {\bf A} and {\bf B}, acceleration between {\bf B} and {\bf C} and non-acceleration between {\bf C} and {\bf D}. These acceleration regimes are shown schematically in Fig.~\ref{fig:Rdiag}.

\subsection{Accelerating weighted networks}\label{subsec:weighted}
Having defined the accelerating unweighted evolving network, we now extend the concept to encompass weighted networks. This is a relatively simple process. The key components in the definition of the network acceleration for unweighted graphs were the rates of node and link addition, both of which are macroscopic properties of the system. Similarly, we can observe the macroscopic weight of the system, denoted $L(t)$, which reflects the total weight of all the links within the network expressed as

\begin{eqnarray}\label{eqn:weights}
L(t)&=&\sum_{i=1}^{M(t)}w_i.
\end{eqnarray}

The total weight of the evolving network at time $t$ is constrained by the total number of possible links, given by $L_{\max}(t) = N(t) (N(t)-1)/2$. 
 

In a similar manner to the unweighted scenario (see Eq.~\ref{eqn:approximate} for comparison), we make use of the approximate derivatives and write
\begin{eqnarray} ~\label{eqn:weightedderivs}
l(t) & = \frac{d L(t)}{d t} & \approx L(t) - L(t-1) \nonumber\\
n(t) & = \frac{d N(t)}{d t} & \approx N(t) - N(t-1).
\end{eqnarray}
For useful comparison between networks, it in important to normalise the weights such that for any link $w\in[0,1]$ otherwise $l(t)$ would vary enormously according to the type of network under consideration\footnote{The method employed with which to perform this normalisation is likely to be situation specific. A simple division by the largest weight in the network might suffice or some cumulative binning process might be appropriate. In either case, it is necessary to take care with respect to the statistical significance of outliers.}
The evolution of these rates for weighted graphs are then used to define velocity and acceleration as

\begin{eqnarray}\label{eqn:ratioweighted}
\tilde{v}(t)&\equiv& \frac{l(t)}{n(t)} \nonumber\\
\tilde{a}(t)&\equiv&\frac{d}{d t}\left(\frac{l(t)}{n(t)}\right). 
\end{eqnarray}

Note that if the weights are restricted to be binary in nature, the above weighted definitions of Eq.~\ref{eqn:ratioweighted} recover the unweighted definitions of Eq.~\ref{eqn:approximate}, i.e., $\tilde{v}(t) \to v(t)$ and $\tilde{a}(t) \to a(t)$ as weights are made binary, which is a desirable feature of any weighted network metric. The above definition is then possibly the most general definition in that it can be applied to both weighted and unweighted, as well as directed or undirected, evolving networks.

\subsection{Stochastic accelerating networks}\label{subsec:stochastic}
The definitons (weighted and unweighted) outlined in this section have been introduced for the scenario whereby the rates of node and link addition are not stochastic. They are, infact, easily applied to stochastic situations. In this case, the corresponding measures would be the expected velocity $\langle v(t) \rangle$ and expected acceleration $\langle a(t)\rangle$. In certain cases, it might be possible to achieve this by simply replacing $n(t)$, $m(t)$ and $l(t)$ by their expectation values although in general this will {\em not} be suitable. For this to be appropriate we require rather specific constraints on the evolution of the system, namely that the rate of node addition $n(t)$ is deterministic and the rate of link (weight) addition is not path dependent. That is, $m(t)$ is not dependent on the number of links added at the last time step (i.e. is independent of $M(t-1)$).

In general, to evaluate the required quantities properly, we must consider all possible contributing trajectories of the system's evolution and their corresponding probabilities. We must incorporate the possibility that the rates of node and link addition at a given time step might not be independent of each other and that their outcomes might also influence the evolution of the network at the next time step. For the unweighted case, this would give
\begin{small}
\begin{eqnarray}
\langle v(t) \rangle &=& \sum_{a,b} P[(m(t)=a)\cap(n(t)=c)]\left(\frac{a}{c}\right)  \nonumber\\
\langle a(t) \rangle &=& \sum_{a,b,c,d} \Bigg( P[(m(t+1)=b)\cap(n(t+1)=d)\cap(m(t)=a)\cap(n(t)=c)] \nonumber\\
{}&{}& \left(\frac{b}{d}-\frac{a}{c}\right) \Bigg).
\end{eqnarray}
\end{small}

\section{Case study: Wikipedia} \label{sec:case}

\begin{figure}[htbp]
\begin{center}
\includegraphics[width=0.8\textwidth]{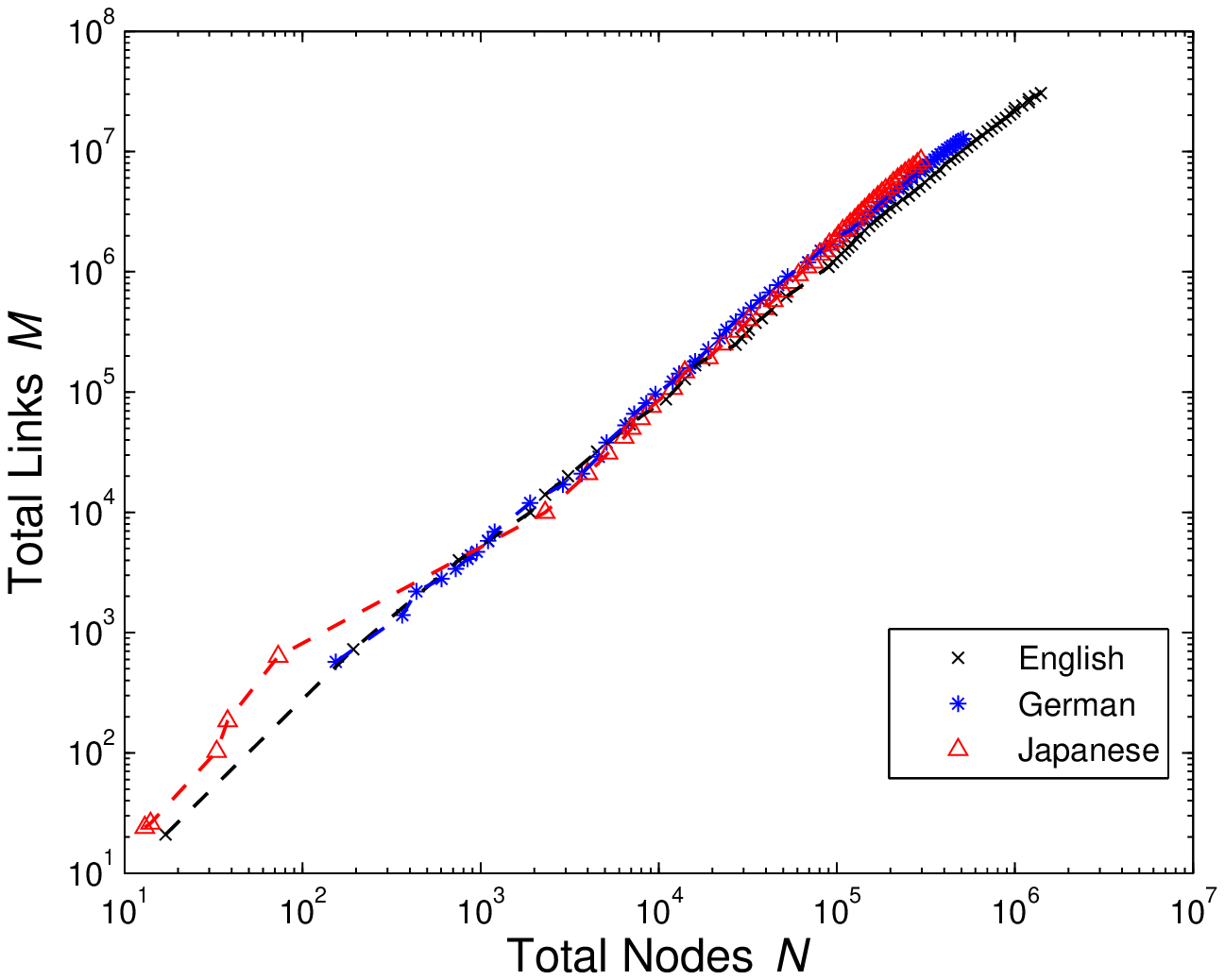}
\includegraphics[width=0.8\textwidth]{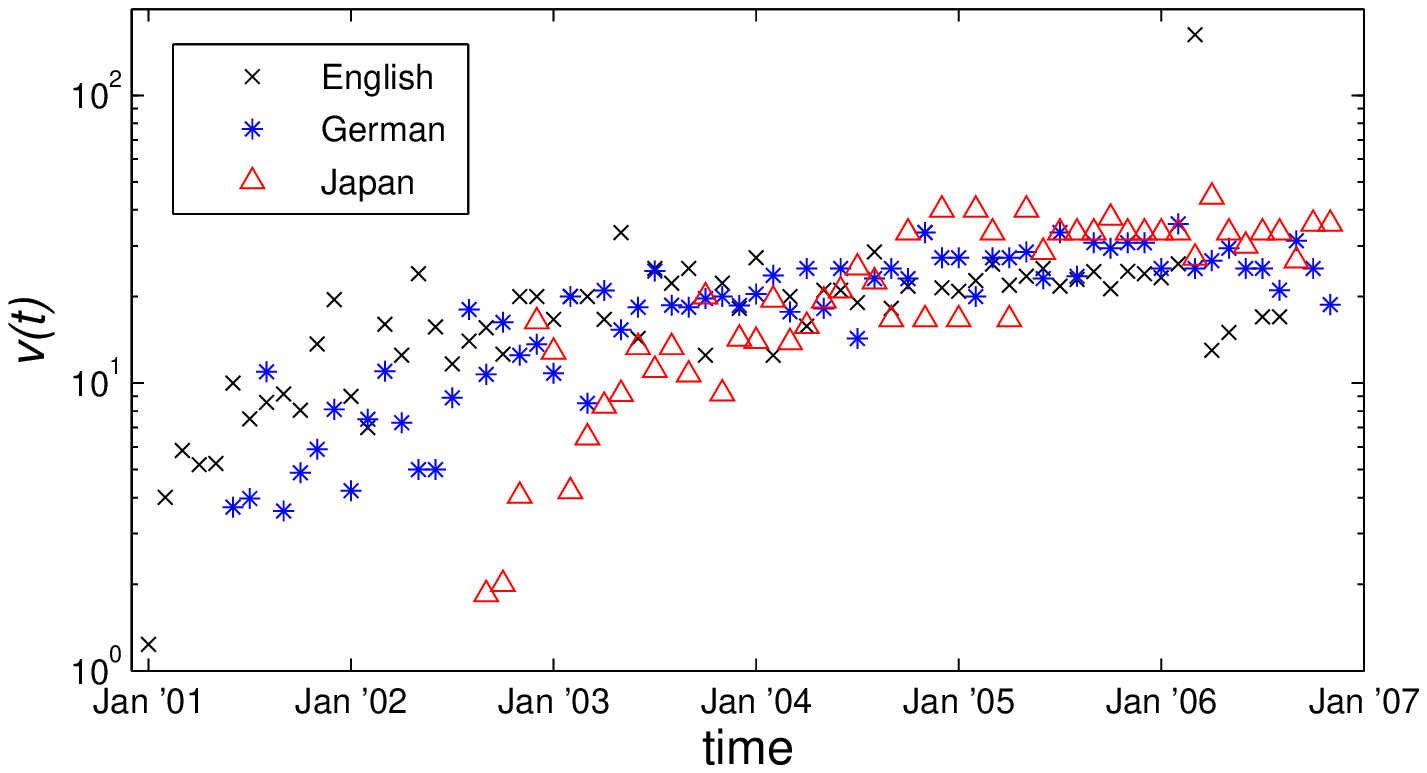}
\caption{ \label{fig:wiki}(Color online) The evolution of the Wikepedia site network for different languages. This comprises articles (nodes) and internal links (links) for the English, German and Japanese sites. The evolution through the macroscopic $M(t)$ and  $N(t)$ phase space of total node and link numbers is shown in the top plot and the network velocity in the lower. }
\end{center}  
\end{figure}

As a simple example of application of the concepts above introduced, we look at the evolution of Wikipedia in three different languages, namely,  in English, German and Japanese. Each of these is a distinct evolving network, such that the nodes correspond to the articles and the links correspond the the links between articles. The data, albeit imprecise,  is available in the public domain \cite{wikidata}. The evolution through the macroscopic $M(t)$ -  $N(t)$ phase space is shown in the upper panel of Fig.~\ref{fig:wiki}. All three networks appear to converge on the same non-accelerating behaviour. 

While the public domain data is somewhat imprecise, the plot of the evolution of the velocity $v(t)$ of the system does indicate that all networks show an initial accelerating trend before non-accelerating behaviour is reached as shown in the lower plot of Fig.~\ref{fig:wiki}. There are no negative velocities as the total number of links and the total number of nodes are increasing in time for all three networks. It is interesting to note that acceleration of the growing Japanese site far exceeded that of the English and German sites. This simple example demonstrates that it is reasonable to consider network velocity and acceleration as time-dependent properties of networks as opposed to considering them as static properties of networks as suggested in \cite{GM}. It is also very straightforward to measure the introduced characteristics, velocity and acceleration, for empirical networks, which further supports their role in network characterization.

\section{Conclusion} \label{sec:conclusion}
We have revisited the framework of network acceleration suggested by Gagen and Mattick \cite{GM}. We have explored the limits of the proposed definition of network acceleration and, based on our findings, have provided an alternative definition for accelerating networks. Perhaps most important is the conceptual difference between the two definitions: the concept of network acceleration as introduced in this paper refers to the properties of the network at a particular moment in time as opposed to an algorithm governing the evolution of the network as suggested in \cite{GM}. In addition to introducing the related concept of network velocity, we have augmented the definition of network acceleration to cover weighted networks as well. As such, the definition put forward in this paper holds for both weighted and unweighted, as well as directed and undirected graphs. We have demonstrated the utility of these concepts by their simple application to study the evolution of Wikipedia in three different languages. While the data obtained from public domain is not very accurate, the obtained results clearly support the conclusion that networks undergo different regimes of acceleration throughout their evolution. Since measurement of the proposed characteristics for empirical networks is very simple, we hope that the measures will find their use in the study of network evolution, in particular as time-stamped network data becomes increasingly available in the future.

\vspace{1cm}

\noindent \textbf{Acknowledgements:} 
D.M.D.S. acknowledges support from the European Union under the MMCOMNET program and J-P.O. by a Wolfson College Junior Research Fellowship (Oxford, U.K.).
\vspace{1cm}



\begin{thebibliography}{99}
\bibitem{mattick:2005} J. S. Mattick and M. J. Gagen. Accelerating networks. {\it Science}, 307:856, 2005.
\bibitem{internet} S. N Dorogovtsev and J. F. F. Mendes. Effect of the accelerating growth of communications networks on their structure. {\it Phys. Rev. E}, 63:2510, 2001.
\bibitem{GM} G.M Gagen and J. S. Mattick. Accelerating, hyperaccelerating and decelerating probabilistic networks. {\it Phys. Rev. E}, 72:16123, 2005.
\bibitem{Sen} P. Sen.  Accelerated growth in outgoing links in evolving networks: Deterministic versus stochastic picture. {\it Phys. Rev. E}, 69:46107, 2004. 
\bibitem{NAF} See ``http://www.physics.ox.ac.uk/users/smithdmd/FDN.pdf" for details.
\bibitem{dorogovtsev} S. N. Dorogovtsev and J. F. F. Mendes. {\it Evolution of Networks: From Biological Nets to the Internet and WWW}. Oxford University Press, 2003.
\bibitem{ER} E. Erd\"os and A. R\'enyi. On random graphs. {\it Publ. Math. Debrecen}, 6:290, 1959.
\bibitem{KrapivskyDeriv} P. L. Krapivsky, S. Redner, and F. Leyvraz. Connectivity of growing networks. {\it Phys Rev. Lett.}, 85:4629, 2000.
\bibitem{DorogovtsevDeriv} S. N. Dorogovtsev, J. F. F. Mendes, and A. N. Samukhin. Structure of growing networks with preferential linking. {\it Phys Rev. Lett.}, 85:4633, 2000.
\bibitem{BarabasiDeriv1} A. L. Barab\'asi and R. Albert. Emergence of scaling in random networks. {\it Science}, 286:509, 1999. 
\bibitem{BarabasiDeriv2} A. L. Barab\'asi,  R. Albert and H. Jeong. Mean-field theory for scale-free random networks. {\it Physica A}, 272:173, 1999.  
\bibitem{Markov} See ``http://mathworld.wolfram.com/MarkovChain.html" for details on the Markov Chain.
\bibitem{wikidata} See ``http://stats.wikimedia.org/EN/Sitemap.htm" for Wikipedia site statistics.
\end{thebibliography}
\end{document}